\documentclass{ptephy_v1}

\usepackage{amssymb}
\usepackage{amsmath}
\usepackage{url}
\definecolor{lightgray}{gray}{0.95}
\graphicspath{{./}}

\begin{document}

\title{Constructing approximate shell-model wavefunctions by eigenvector continuation}


\author{Sota Yoshida}
\affil{Liberal~and~General~Education~Center,~Institute~for~Promotion~of~Higher~Academic~Education, Utsunomiya~University,~Mine,~Utsunomiya,Tochigi~321-8505,~Japan\email{syoshida@cc.utsunomiya-u.ac.jp}}
\author[2]{Noritaka Shimizu}
\affil{Center for Nuclear Study, the University of Tokyo,~Hongo,~Bunkyo-ku,~Tokyo~113-0033,~Japan}

\begin{abstract}%
Shell-model calculations play a key role in elucidating various properties of nuclei.
In general, those studies require a huge number of calculations to be repeated for parameter calibration and quantifying uncertainties.
To reduce the computational burden, we propose a new workflow of shell-model calculations using a method called eigenvector continuation (EC).
It enables us to efficiently approximate the eigenpairs under a given Hamiltonian by previously sampled eigenvectors. We demonstrate the validity of EC as an emulator of the valence shell-model, including first application of EC to electromagnetic transition matrix elements.
Furthermore, we propose a new usage of EC: preprocessing, in which we start the Lanczos iterations from the approximate eigenvectors, and demonstrate that this can accelerate subsequent research cycles.
With the aid of the EC, the eigenvectors obtained during the parameter optimization are not necessarily to be discarded, even if their eigenvalues are far from the experimental data.
Those eigenvectors can become accumulated knowledge.
\end{abstract}

\subjectindex{D1}

\maketitle

\section{Introduction} \label{sec:Intro}

The configuration interaction method plays a key role to study various properties of many-fermion systems. Nuclei are no exception; the nuclear shell model is one of the most powerful approaches to a microscopic understanding of static properties of nuclei~\cite{Brown_Rev,CaurierRev,RMP_Otsuka}.
Its application covers a wide range of the nuclear chart, and it also provides a bridge between {\it ab initio} and phenomenological studies with the development in many-body methods to derive the shell-model Hamiltonians: Over the past decades, we have witnessed much progress in this direction with a nuclear force within chiral effective field theory (chiral EFT)~\cite{EGMrev,EMrev}.
A representative and up-to-date example of the many-body method is the valence-space in-medium similarity renormalization group (VS-IMSRG)~\cite{VS-IMSRG1,VS-IMSRG2,VS-IMSRG3,Stroberg_rev19,VS-IMSRG_Miyagi}, which has enlarged the reach of shell-model studies with non-empirical interactions to nearly 700 isotopes~\cite{Stroberg2021}.
Simultaneously, there have been much progress by other approaches such as coupled-cluster effective interaction~\cite{CCEI1,CCEI2}, no-core shell-model and Okubo-Lee-Suzuki transformation~\cite{EI_NCSM1,EI_NCSM2}, shell-model coupled cluster~\cite{SMCC1,SMCC2}, and so on.

\begin{figure*}[t]
\centering{
\includegraphics[width=15cm]{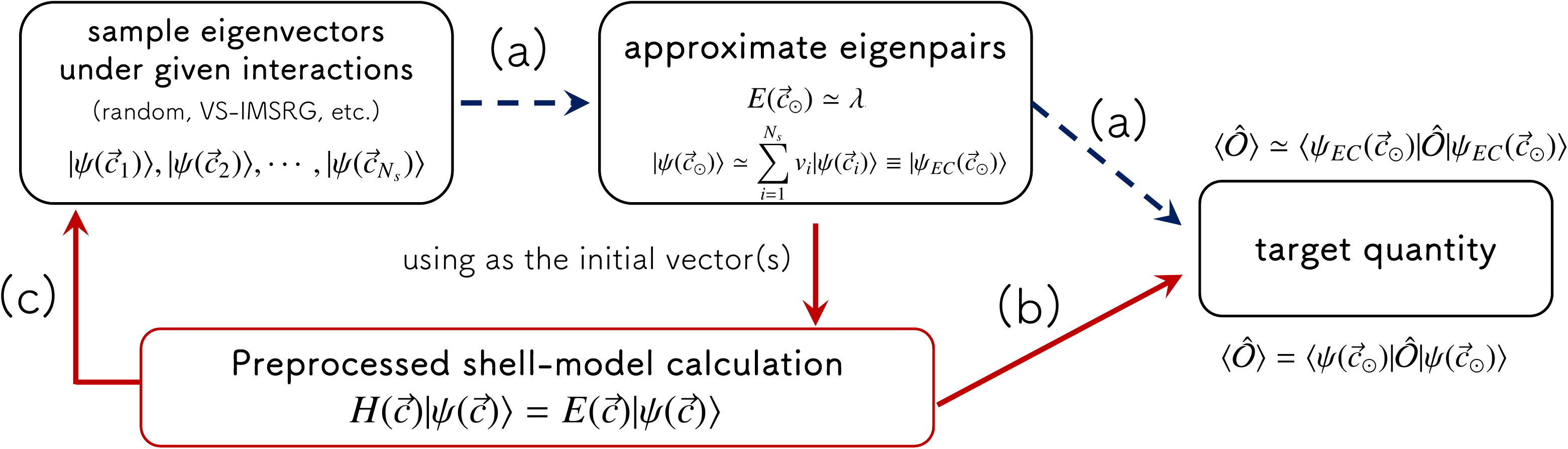}
\caption{A new workflow of the shell-model calculations:
(a) EC can be used as an emulator of shell-model calculations, (b) preprocessed exact calculations for target quantities, (c) making another sample eigenvector to improve the accuracy of the emulator.
See main text for more details.
\label{pic:flowchart}}
}
\end{figure*}

In practice, however, it is still necessary to repeat a huge number of shell-model calculations to optimize the effective interactions, quantify uncertainties, extract physical intuitions by comparison with experimental observations, test the validity of many-body methods such as VS-IMSRG, and so on.
One of the aims of this work is to propose a new workflow of shell-model calculations, i.e., how such a research procedure using shell-model calculations can be accelerated.
In Fig.~\ref{pic:flowchart}, we show a schematic picture of the workflow.
As will be detailed in the followings, the key ingredient is the eigenvector continuation (EC).

While much efforts have been spent on calibrating input parameters for nuclear models (e.g.,
low-energy constants in chiral EFT and phenomenological shell-model interactions),
one still needs expensive numerical samplings in a high-dimensional parameter space to find a reasonable range of those parameters and to quantify the associated uncertainties in the parameters and the target observables for a deeper understanding of properties of nuclei.

Recently, the importance of the uncertainty quantification (UQ) has been widely pointed out in
various contexts, such as parameter calibration of the chiral EFT potentials~\cite{Carlsson2016PRX,Melendez2017,Wesolowski2019} and nuclear observables~\cite{UQ_DFT,UQ_LDM,UQ_NM_1,UQ_NM_2,UQ_reaction}.
This also applies to the shell model, but the UQ for the valence shell-model studies are limited up to the $sd$ shell~\cite{SY_UQ,Fox_UQ}.
This is due to the rapid growth in the size of matrices to be diagonalized as the number of valance nucleons or the size of model space (called valence space too) increases.
To reduce the computational cost for UQ, an approximation method using the principal component analysis was proposed in Ref.~\cite{Fox_UQ}.
However, one still needs additional efforts and/or efficient methods for parameter calibration and UQ to enlarge the scope to a heavier or neutron/proton-rich region.

The paper is organized as follows. In Section~\ref{sec:Formalism}, we explain some basics of shell-model calculations and the eigenvector continuation.
We show a couple of applications of EC as an emulator and as a preprocessing in Section~\ref{sec:Results}, and conclude and give an outlook in Section~\ref{sec:Summary}.
Technical details on formulations, codes, and results are given in Supplementary Material~\cite{supple} for reproducibility.

\section{Formalism \label{sec:Formalism}}

\subsection{Nuclear shell model}

In shell-model calculations, the problem of interest is to solve the Sch\"odinger equation under a given effective interaction for valence nucleons, which is dependent on some parameter $\vec{c}$:
\begin{align}
H(\vec{c}) | \psi (\vec{c}) \rangle &= E(\vec{c}) | \psi (\vec{c}) \rangle. \label{eq:SchEq}
\end{align}
The typical parametrization of the shell-model Hamiltonian is the following:
\begin{align}
H = H^{(1)} +  H^{(2)} = \sum_{ac} h^{(1)}_{ac} c^\dag_a c_c + \frac{1}{4} \sum_{abcd} h^{(2)}_{abcd} c^\dag_a c^\dag_b  c_d c_c, \label{eq:SMHamil}
\end{align}
where $c^\dag_a$ and $c_c$ denote a creation and annihilation operator on the single-particle state $a$ and $c$, respectively.
One can naturally extend the above to include the three-body (and higher many-body) term, but we will not go into the details of three-body forces in this work.

The eigenvectors, called wave functions too, are expressed as a superposition of many-body configurations represented by Slater determinants.
We employ the so-called $M$-scheme basis to express the wave function.
This means that the single particle states, $a,b,c,d$ in Eq.~\eqref{eq:SMHamil}, are specified by its harmonic oscillator quanta, $n,l,j$ (and the $z$ component of angular momentum $m_z$ and of isospin $t_z$, if needed).
Under this, the two-body part of shell-model Hamiltonian, $H^{(2)}$, can be rewritten in a more explicit form:
\begin{align}
& H^{(2)} = \frac{1}{4}\sum_{abcdJM} 
\mathcal{N}_{ab}(J)\mathcal{N}_{cd}(J)  A^\dag(ab;JM) A(cd;JM) V_J(abcd), \label{eq:H2} \\
& \mathcal{N}_{ab}(J) = \left[ (1+\delta_{ab}) \right]^{1/2}, 
\mathcal{N}_{cd}(J) = \left[ (1+\delta_{cd}) \right]^{1/2}, \\ 
& A^\dag(ab;JM)  = 
\sum_{m_a,m_b} (j_am_aj_bm_b|JM)  c^\dag_{j_am_a}c^\dag_{j_bm_b}  \label{eq:Ad} \\
& A(cd;JM)  = \sum_{m_c,m_d}  (j_cm_cj_dm_d|JM) c_{j_dm_d}c_{j_cm_c} \label{eq:A}
\end{align}
where the label of single particle states, $a,b,c,d$, denote the quanta $\{n,l,j,t_z\}$, $(j_am_aj_bm_b|JM),(j_cm_cj_dm_d|JM)$ are Clebsch-Gordan coefficients,
and $J$ and $M$ are coupled angular momentum and its $z$ component.
Due to the symmetries of the nuclear force, the number of non-zero $V_J(abcd)$ is typically on the order of tens or thousands, and can be further reduced by a factor of about two when assuming the isospin symmetry of the effective interaction.

Usually, the diagonal part of $h^{(1)}$ ($c=a$) is called single-particle energies (SPEs), and $V_J(abcd)$ in Eq.~\eqref{eq:H2} is called two-body matrix elements (TBMEs).
These SPEs and TBMEs are given in the interaction file for shell-model codes.
Note that, in some codes, the so-called isospin formalism $V_{JT}(abcd)$ is used for interactions with the isospin symmetry.

A typical recipe to determine these SPEs and TBMEs is the following:
(i) making an initial guess in some way, (G-matrix, VS-IMSRG, etc.), and
(ii) modifications of the parameters so as to minimize the chi-square deviation from the selected experimental data.
There have been many previous works adopting the above recipe~\cite{USDB,KB3G,GXPF,GXPF1,GXPF1A,JUN45,USDI}.
One of the most successful and well-known examples is the USDB interaction for the $sd$-shell nuclei~\cite{USDB}.
The USDB gives the root-mean-square deviation $126$ keV for 380 energy data in the mass region $A=16-40$.

Once the interaction is given and the model space is specified, the main task of shell-model codes is diagonalizing the shell-model Hamiltonian under the basis states, which are now in the $M$-scheme. The matrices are in general very sparse, but the total number of non-zero matrix elements is too large to be stored on memory.
For this reason, the diagonalization is typically done by means of the Lanczos method~\cite{Lanczos}.

\subsection{Implementation: ShellModel.jl}

Although many CI codes for many-nucleon systems written in Fortran are already available (e.g., NuShellX~\cite{NuShellX}, BIGSTICK~\cite{BIGSTICK,BIGSTICK2}, ANTOINE~\cite{ANTOINE}, MFDn~\cite{MFDn1,MFDn2,MFDn3}, MSHELL64~\cite{MSHELL64} and KSHELL~\cite{KSHELL1,KSHELL2}), we developed a new code ShellModel.jl~\cite{SY_Github}
written in the Julia language~\cite{Julia,Julia2} for efficient samplings of shell-model results and demonstrating the benefit of the methodology in the present paper.
The Julia language became highly popular in the scientific computing community
because it provides interactive and dynamic behavior and high-productivity like Python, high-performance, and flexibility to combine other modules (e.g., visualization, neural networks, etc.) with its own package management system.

Most of the functions in ShellModel.jl is originally implemented combining other Julia packages, but we referred to MSHELL64 and KSHELL for implementation of some methods.
More precisely, we implemented the followings: the Thick-Restart Lanczos method, block family of the Lanczos methods, and the double Lanczos method~\cite{MSHELL64,Mizusaki2010} for the projection to a specified total angular momentum $J$.
While the KSHELL code adopts the so-called on-the-fly generation of the matrix elements for the  matrix-vector product, ShellModel.jl stores matrix elements rather explicitly on memory to balance the speed and moderate memory usage.
The proton-neutron interaction is stored in the form of ``one-body jumps"~\cite{BIGSTICK,BIGSTICK2}.
The current version of ShellModel.jl is optimized to run on a relatively smaller environment such as a laptop.
The main codes, inputs, and sample scripts used in this study are available on the GitHub repository~\cite{SY_Github} for reproducibility.
We hope that it will facilitate other future studies.

\subsection{Eigenvector continuation: an efficient emulator and a preprocessing for the nuclear shell model}

The eigenvector continuation (EC), which was proposed in Ref.~\cite{EC_Frame}, has been widely used as an efficient emulator of nuclear many-body methods~\cite{EC_CC,EC_Scattering,EC_Scattering2,EC_Scattering3,EC_NCSM,EC_RMat,EC_LEC_NCSM}
and as a resummation method~\cite{EC_BMBPT1,EC_BMBPT2}.
In the following, we give a brief overview of the eigenvector continuation, and then explain how to apply it to shell-model calculations.

Suppose that the eigenvectors for Eq.~\eqref{eq:SchEq} have been already obtained at $N_s$ different parameters $|\psi(\vec{c}_1)\rangle,|\psi(\vec{c}_2)\rangle,\ldots,|\psi(\vec{c}_{N_s})\rangle$.
These eigenvectors will be hereafter referred to as {\it samples}.
In such a case, the eigenpairs under a given Hamiltonian $H (\vec{c}_\odot)$ are approximated by solving the following generalized eigenvalue problem in the subspace spanned by the samples:
\begin{align}
\tilde{H} \vec{v} &= \lambda N \vec{v}, \label{eq:GEV}\\
\tilde{H}_{i,j} & = \langle \psi (\vec{c}_i) | H (\vec{c}_\odot) | \psi (\vec{c}_j) \rangle, \label{eq:Hmat}\\
N_{i,j} & = \langle \psi (\vec{c}_i)  | \psi (\vec{c}_j) \rangle  \label{eq:Nmat}.
\end{align}
Then, the original eigenpairs can be approximated as 
\begin{align}
E(\vec{c}_\odot)  &\simeq \lambda, \label{eq:lambdaEC} \\
|\psi(\vec{c}_\odot) \rangle & \simeq \sum^{N_s}_{i=1} v_i |\psi(\vec{c}_i) \rangle \equiv | \psi_{EC}(\vec{c}_\odot)  \rangle.
\label{eq:psiEC}
\end{align}

The problem is reduced from the diagonalization of a sparse Hamiltonian matrix $H$ with size of $M$-scheme dimension to the $\tilde{H}$ of a dense matrix of size $N_s$, which is typically on the order of tens or hundreds.
This might significantly reduce the computational cost compared to the original problem.
Note that it is straightforward to extend Eqs.~\eqref{eq:GEV}--\eqref{eq:psiEC} to include excited states.
With the approximated wave functions, one can approximate the expectation values
for a target observable:
\begin{align}
\langle \hat{\mathcal{O}} \rangle &\simeq \langle \psi_{EC}(\vec{c}_\odot) |\hat{\mathcal{O}}| \psi_{EC}(\vec{c}_\odot) \rangle,
\label{eq:obs} 
\end{align}
where the operator $\hat{O}$ can be e.g., electromagnetic transition operators.

The eigenvector continuation has been well known as the Rayleigh-Ritz method in (applied) mathematics, and introduced to or re-evaluated in the nuclear physics community recently.
However, its properties are still under consideration~\cite{EC_conv}, and the previous works in nuclear physics have proved for the first time its efficiency for large-scale many-body problems and enlightened the possibility to enlarge the scope of microscopic studies to a heavier and/or more exotic region of the nuclear landscape.

For solving the EC problems, the most time-consuming part is, in general, to evaluate the expectation value of Hamiltonians, i.e., $\tilde{H}$ in Eq.~\eqref{eq:Hmat}.
However, the computational cost for evaluating $\tilde{H}$ can be somewhat alleviated in shell-model calculations.
This is because of the fact that the expectation values of shell-model Hamiltonians can be factorized out by the SPEs and TBMEs: 
\begin{align}
\tilde{H}_{i,j} = 
\sum_{k}  h^{(1)}_{k} \times \overline{\mathrm{OBTD}}_k
+
\sum_{k} V_J(abcd)_k \times \overline{\mathrm{TBTD}}_k ,
\label{eq:TDfac}
\end{align}
where $k$ denotes the labels of one- and two-body interactions in a given shell-model Hamiltonian $H(\vec{c}_\odot)$, and the OBTD and TBTD are the abbreviations of one- and two-body transition densities, respectively.
The concrete expressions for the $\overline{\mathrm{OBTD}}_k$ and $\overline{\mathrm{TBTD}}_k$ are summarized in Sec.~A of Supplementary Material~\cite{supple}.
Once we evaluate the transition densities in Eq.~\eqref{eq:TDfac} for arbitrary two sample eigenvectors, one can significantly reduce the computational cost to evaluate $\tilde{H}$ for different parameters, because these can be evaluated simply by the dot product of the parameters and the transition densities, which are independent on the parameter values.

Although the number of required samples depends on the problem and desired accuracy,
the cost for solving generalized eigenvalue problem in Eq.~\eqref{eq:GEV} is typically negligible compared to that of the original eigenvalue problem.
Hence, the eigenvector continuation provides an efficient emulator of the shell-model calculation as well as other nuclear models discussed in the previous works~\cite{EC_CC,EC_Scattering,EC_Scattering,EC_Scattering2,EC_Scattering3,EC_NCSM,EC_RMat,EC_LEC_NCSM}

In addition to its role as an emulator, EC can also be used as a preprocessing method. 
In the Lanczos method, one usually starts with the Lanczos iterations from a random vector, when one has no prior information about the starting vector.
The convergence of the Lanczos iterations should be accelerated by starting from a good initial guess of the eigenvectors.
The approximate eigenvectors constructed by EC would make it, and it will be also helpful to increase the sample size for further improvement of the accuracy the EC emulator.

\section{Results}\label{sec:Results}

In this section, we present the main results of this study.
For demonstration purposes, we restrict ourselves to mainly consider the $sd$ shell on top of ${}^{16}$O core, i.e., the model space consists of $1s_{1/2},0d_{3/2},0d_{5/2}$ orbits (12 single-particle states in total) for both protons and neutrons.
Since we consider only the positive parity states, the plus sign on the total angular momentum $J$, e.g., $0^+, 2^+$, will be omitted where appropriate.

In Sec.~\ref{sec:SMjl}, we briefly describe our procedures to prepare the sample eigenvectors.
Next, the efficiency of the EC emulator is demonstrated in Sec.~\ref{sec:En} and Sec.~\ref{sec:MCMC}.
We also make some remarks on the accuracy of approximated eigenvectors in terms of magnetic dipole moments and electric quadrupole moments. In Sec.~\ref{sec:preprocess}, we explain the preprocessing for the Lanczos method and discuss its possible accelerating effect on the convergence. 
In Sec.~\ref{sec:ReEM}, we re-examine the accuracy of EC approximate wavefunctions utilizing the preprocessing.
Finally, we will mention the extensibility of the proposed method for larger systems in Sec.~\ref{sec:exten}.

\subsection{Calculations of $sd$-shell nuclei with ShellModel.jl}\label{sec:SMjl}

Our new shell-model code, ShellModel.jl, is designed to make sample eigenvectors efficiently and to demonstrate the efficiency of the emulator and the preprocessing with EC.
With ShellModel.jl, the execution time to evaluate 10 lowest states of ${}^{28}$Si, which has the largest $M$-scheme dimension in the full $sd$-shell space (93,710 for $M=0$), is about 3 seconds on a MacBook Air (2020, Apple M1).
We provide the sample codes in the GitHub repository~\cite{SY_Github}, and further instructions for the sample codes are given in Sec.~B of Supplementary Material~\cite{supple}.

To make the sample eigenvectors, we generated 50 random $sd$-shell interactions by adding gaussian random values with the standard deviation $\sigma_\mathrm{int.}=1.0$ MeV to the USDB interaction~\cite{USDB}.
Hereafter, the terms ``random interaction" and ``sample" denote those by $\sigma_\mathrm{int.}=1$ unless otherwise mentioned. In the next section, we also discuss the results with $\sigma_\mathrm{int.}=3$ and samples utilizing Latin Hypercube Sampling~\cite{LHS}.
The USDB is a phenomenological $sd$-shell interaction constructed by $G$-matrix~\cite{MortenG} and the chi-square fit using 608 data in 77 nuclei.
In accordance with the USDB interaction having the isospin symmetry,
the Hamiltonians are defined in the 66-dimensional parameter space (3 for SPEs and 63 for TBMEs).
Next, we calculated the five lowest eigenvectors with $J=0,1,2,3,4$ (for even nuclei) and $J=1/2,3/2,5/2,7/2,9/2$ (for odd) under the 50 random interactions, i.e. sampling 250 eigenvectors in total for each $J$.

\subsection{Approximate eigenpairs by the eigenvector continuation\label{sec:En}}

\begin{figure*}[h]
\centering{
\includegraphics[width=7.5cm]{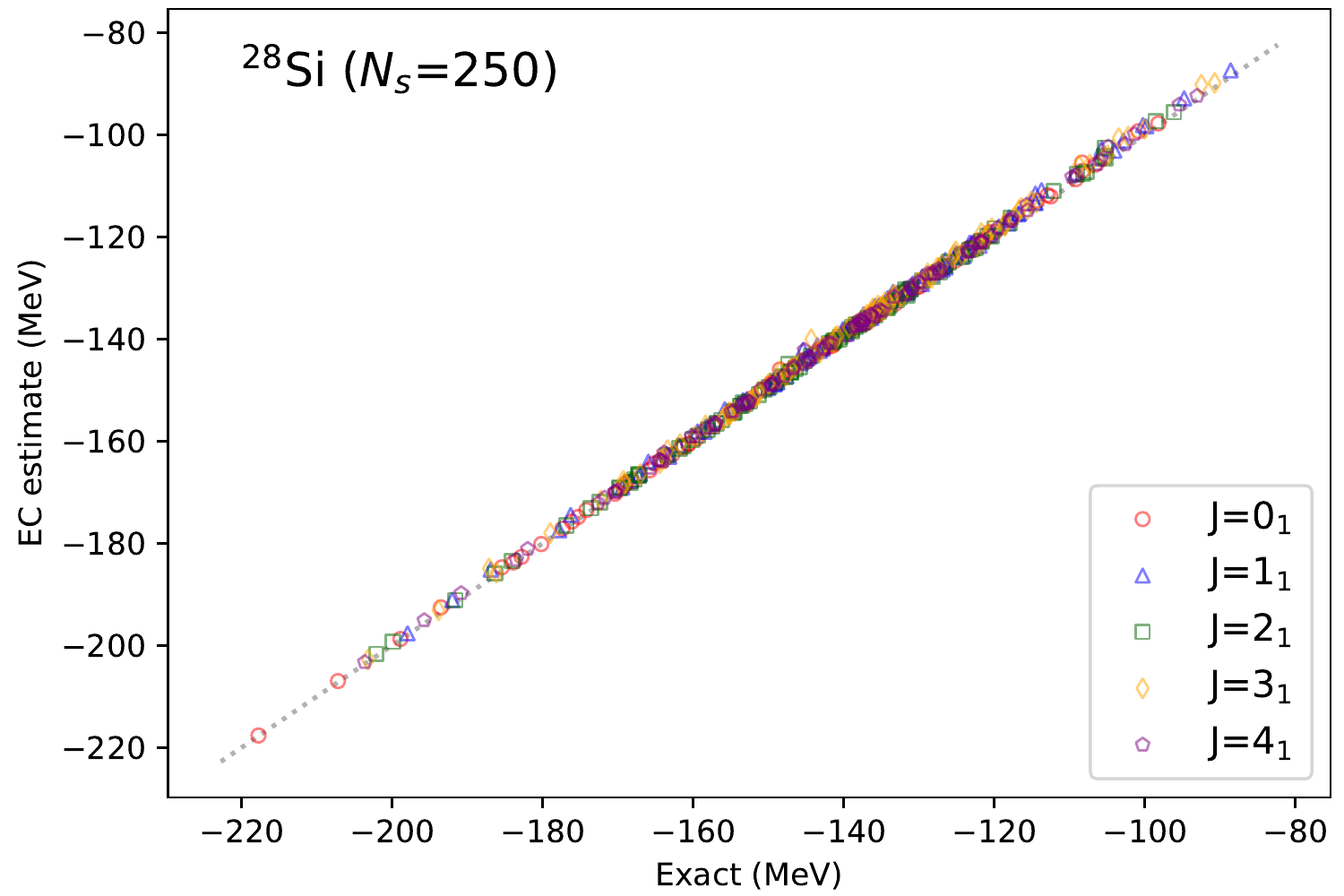}
\includegraphics[width=7.5cm]{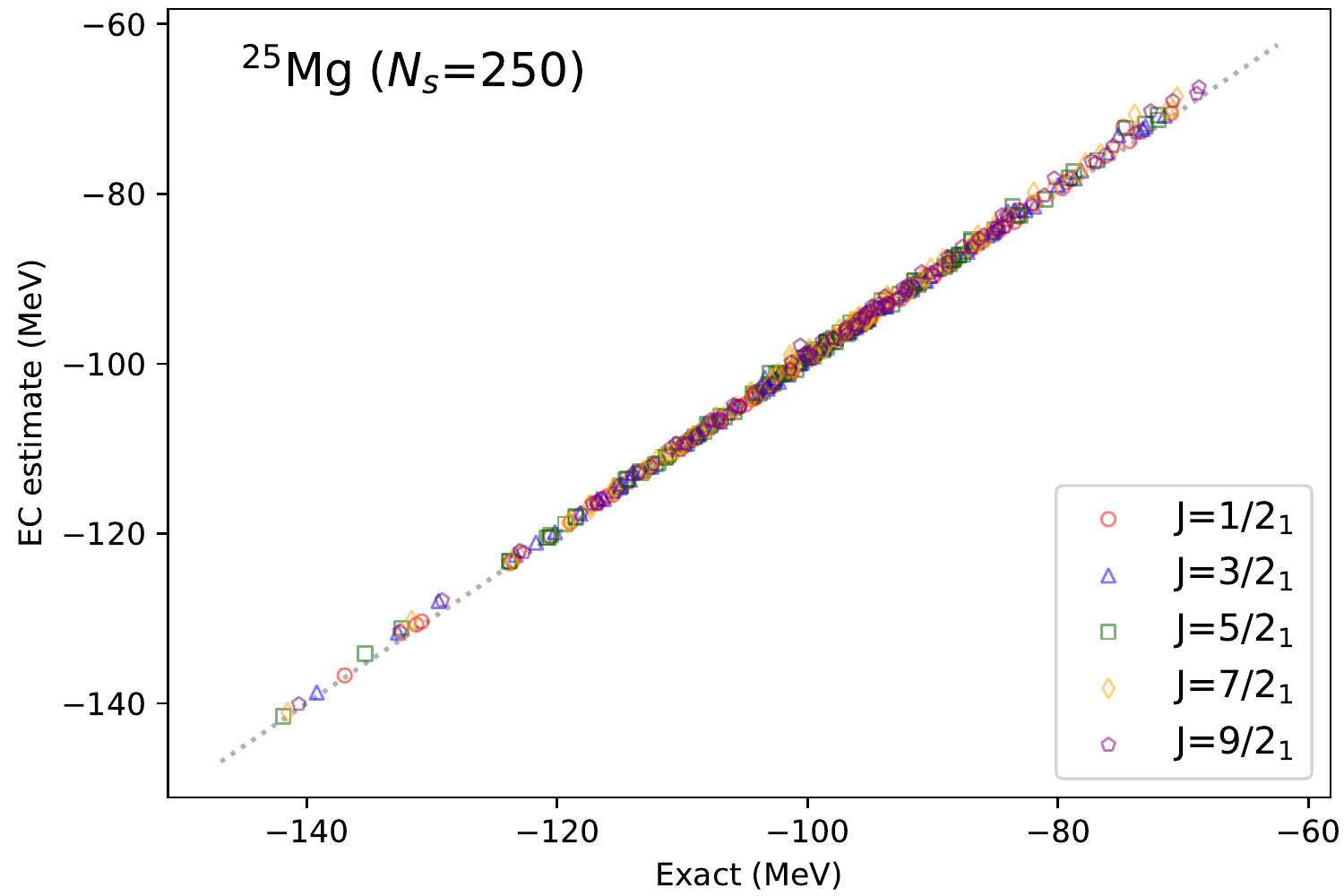}
\caption{Comparison between the exact energy eigenvalues and the EC estimates for 100 random interactions.
The upper and lower panels show the five yrast states of ${}^{28}$Si and ${}^{25}$Mg, respectively.\label{pic:Si28Mg25}}}
\end{figure*}

In what follows, we show the results of approximate eigenpairs by EC.
Using the samples discussed in the previous subsection, we solved Eq.~\eqref{eq:GEV} to estimate the approximate eigenvalues for other 100 random interactions, which were made in the same way as above, and compare them to the exact values.
In the followings, the 100 random interactions are referred to as the validation set.

In Fig.~\ref{pic:Si28Mg25}, we show the EC estimates of five yrast states for ${}^{28}$Si $(J=0,1,2,3,4)$ and ${}^{25}$Mg $(J=1/2,3/2,5/2,7/2,9/2)$ in comparison with the exact values.
One can see from Fig.~\ref{pic:Si28Mg25} that the symbols are on the diagonal (dotted line) while the absolute values of energy under the validation set spread over the relatively large range.

In Table~\ref{tab:err}, we summarize the typical size of two types of errors by EC estimates compared to the exact results.
One is the relative error of absolute energy values, and the other one is error in terms of the excitation energies:
\begin{align}
\mathrm{relative\ error} (\%) &\equiv 100 \left| \frac{ E_\mathrm{exact} - E_\mathrm{EC} }{E_\mathrm{exact}} \right|, \label{eq:relerr}\\
\mathrm{ex.\ error\ (MeV)} & \equiv 
|E^\mathrm{ex.}_\mathrm{exact} - E^\mathrm{ex.}_\mathrm{EC}|.
\end{align}
Since excitation energies themselves are dependent on the level ordering,
we restrict ourselves to measure excitation energies from the $0^+$ $(1/2^+)$ state for even (odd) nuclei.

As shown in Table~\ref{tab:err}, the typical size of the relative errors is less than $1\%$, when all the $N_s=250$ samples for each $J$ is used.
Here we chose the four nuclei in the middle of $sd$ shell (${}^{28}$Si, ${}^{26}$Al, ${}^{25}$Mg, and ${}^{24}$Mg) as examples.
We found that the relative errors for the odd-odd nucleus, ${}^{26}$Al, are worse than those for the others.

In the context of nuclear structure,
the excitation energy may be of interest rather than the absolute value of the energy.
As shown in Table~\ref{tab:err}, the typical size of the error lies on the order of a few hundred keV or 1 MeV.
The errors for excitation energies show relatively slower convergence to the exact ones compared to their absolute energies. This can be general tendency because the excitation energies are sensitive to the relative convergence speed of the EC wavefunctions.
The origin of this tendency will be examined in the Sec.~\ref{sec:exten}.

To show the dependence on the sampling procedure, we summarized the results for five different $N_s$ in Table~\ref{tab:err}. The sample size $N_s$ is the product of the number of random interactions and the number of excited states used as the sample eigenvectors for EC, and some cases are marked with an asterisk to indicate that the detailed conditions differ from the others.
For example, the $N_s=50^*$ case means that only the first and second lowest states under the first half of the 50 random interactions are used, while the $N_s=250^*$ ($\sigma_\mathrm{int.}=3$) case corresponds to the result using the samples, where the size of Gaussian noise $\sigma_\mathrm{int.}$ increased from $1$ to $3$.
The eigenvalues of the five yrast states of ${}^{28}$Si under the $N_s=250^*$ ($\sigma_\mathrm{int.}=3$) spread over a wide range of $-350$ to $-50$ MeV,
but the relative errors are still $\sim3\%$.
This indicates that one can make use of the eigenvector continuation for rough parameter optimization even if the prior knowledge of the parameter range is poor.

Besides the random samples generated with Gaussian noises,
we tested the dependence on the sampling procedure by utilizing Latin Hypercube Sampling (LHS)~\cite{LHS}.
To compare with gaussian random samples, we fix the lattice size $L$ for LHS to be 2, i.e. to cover 2$\sigma$ ($95\%$) domain with $\sigma_\mathrm{int.}=1$ case.
As shown in Table~\ref{tab:err}, the result of $N_s=250^*$(LHS) samples show almost the same accuracy with the $N_s=250$ case.
One possible advantage of using LHS for EC is that it provides numerical stability for generalized eigenvalue problems, since the distance between each LHS sample tends to be larger than that for the Gaussian case.
In this study, the generalized eigenvalue problem, Eq.~\eqref{eq:GEV} is not ill-conditioned with the help of relatively high dimension, $D=66$.
Hence, we mainly focus on the results with random samples generated with gaussian noise for brevity of the following discussions.

\begin{table}
\centering{
\caption{Average size of two errors by EC estimates for the five yrast states of the four $sd$-shell nuclei: One is the relative error ($\%$) of absolute energies,
and the other one is error of excitation energies. 
The sample size $N_s$ means the product of the number of random interactions and the number of excited states used as the {\it sample} eigenvectors in Eqs.~(\ref{eq:Hmat}-\ref{eq:Nmat}).
The $N_s=250^*$ with $\sigma_\mathrm{int.}=3$ means that the standard deviation to generate the random interactions is increased from the default value $\sigma_\mathrm{int.}=1$,
and the $N_s=250^*$ (LHS, $L=2$) corresponds to the result using Latin Hypercube Sampling (LHS).
\label{tab:err}}
\begin{tabular}{lrrrrrrrrr}
\\
\hline\hline
$N_s$ &  
 \multicolumn{4}{l}{relative error $(\%)$} & 
 &
 \multicolumn{4}{l}{ex. error (MeV) }\\
\cline{2-5}\cline{7-10}s
($\#$ interaction $\times$ $\#$ states)  & ${}^{28}$Si & ${}^{26}$Al & ${}^{25}$Mg & ${}^{24}$Mg 
&
&
 ${}^{28}$Si & ${}^{26}$Al & ${}^{25}$Mg & ${}^{24}$Mg 
\\
\hline
50 ($50\times1$)    &  1.4 & 2.1 & 1.8 & 1.3 & & 0.66 &  1.22  &  0.62 & 0.65\\
$50^*$ ($25\times2$)  & 1.8 & 2.3 & 2.1 & 1.7 & &0.82 &  1.16  & 0.61 & 0.97\\
150 ($50\times3$)  & 0.9 & 1.2 & 1.1 & 0.7 & & 0.44 &  0.85  & 0.42 & 0.62\\
250 ($50\times5$)  &  0.7 & 0.9 & 0.8 & 0.5 & & 0.39 & 0.70 & 0.37 & 0.51 \\
$250^*$ ($50\times5$; $\sigma_\mathrm{int.}=3$)  & 2.8 & 3.3 & 3.1 & 2.3 & & 1.35 & 2.35 & 1.09 & 1.96\\
$250^*$ ($50\times5$; LHS, $L=2$)  & 0.8 & 1.0 & 0.9 & 0.6 & & 0.47 & 0.73 & 0.40 & 0.57\\
\hline\hline
\end{tabular}
}
\end{table}

From the results above, one can see that there are two strategies to improve the accuracy of the EC estimates.
One is to sample more excited states; the $N_s=250$ case using the five lowest states as the samples improves the accuracy by a factor of two in terms of the relative errors from the $N_s=50$ case.
The other one is to increase the sample points in the parameter space.
Looking at the results of the $N_s=50$ and $N_s=50^*$ cases,
we expect that the latter one is more beneficial unless we change the number of states of interest.
On the other hand, the former one is relatively easy in terms of computational costs.
Since the efficiency may depend on the sampling procedure and the distribution of eigenvalues, how to select the strategy and how many states to be sampled will be determined by the trade-off between the costs and desired accuracy.

Note that one can receive benefit from the factorization of transition densities in Eq.~\eqref{eq:TDfac} during the sampling procedure; to increase the samples,
all one needs is evaluating transition densities among already sampled eigenvectors and the new sample.

The execution time to calculate approximate eigenvalues for a random $sd$-shell interaction
is about a few tens of milliseconds in our settings.
It significantly facilitates the sampling procedures for the parameter optimization or uncertainty quantification.

\subsection{Uncertainty quantification\label{sec:MCMC}}

In this subsection, we consider the Monte Carlo sampling over the parameter space utilizing the EC emulator.
The following analysis is by no means an exhaustive study of quantifying uncertainties in the $sd$-shell effective interactions, but it is meant to illustrate the usefulness of the emulator.

We simplify the problem by picking up twelve energy data of ${}^{28}$Si from those used to construct the USDB~\cite{USDB}: $0^+_{1-2}, 1^+_{1-2}, 2^+_{1}, 3^+_{1-4}$, $4^+_{1-3}$.
This is apparently too small data set to constrain the 66 parameters in the $sd$-shell interaction (in the isospin formalism), but enough for the demonstration purpose.

For this small data set, we consider to get samples obeying a certain posterior distribution.
Let the log-likelihood be 
\begin{align}
\log L(\vec{c}) = -\frac{1}{N} \sum^{N}_{i=1} \frac{ (E_{\mathrm{EC},i}(\vec{c}) - E_{\mathrm{Exp.},i})^2}{2\sigma^2_{\mathrm{err},i}},
\label{eq:llh}
\end{align}
where $E_{\mathrm{EC},i}(\vec{c})$, $E_{\mathrm{Exp.},i}$, and $\sigma_{\mathrm{err},i}$ are the EC estimate of energy, its corresponding experimental value, and the expected typical error of the calculation.
Since the errors stemming from EC are dominant, we fixed the $\sigma_{\mathrm{err},i}$ as follows
\begin{align}
\sigma_{\mathrm{err},i}^2 = \sigma_\mathrm{EC,typ.}^2 +\sigma^2_{\mathrm{EC},i},
\end{align}
where the first term is fixed as $1.0$ MeV, which is typical error in order of $1\%$ of the energy eigenvalues, and the second term is small correction with $i$ dependence, which is taken from the difference of $E_\mathrm{EC,i}$ with $ N_s=250$ under the USDB interaction from the corresponding exact value.
Since the number of data is less than the number of parameters, we introduced the independent Gaussian prior, which is equivalent to the L2 regularization, to avoid highly multi-modal structure of the log-likelihood in the parameter space. The log-prior is defined as
\begin{align}
\log Pr(\vec{c}) = -\frac{\Lambda}{2} || H(\vec{c})-H(\vec{c}_\mathrm{ref.}) ||_2.
\label{eq:prior}
\end{align}
Here $||\cdot||_2$ is the L2 norm, i.e., the sum of squared deviation between a parameter $\vec{c}$ and the reference value $\vec{c}_\mathrm{ref.}$, which is now the USDB.
The choice of variance of the Gaussian prior $\Lambda$ is non-trivial, but we fixed $\Lambda=10$ to simplify the discussions.

\begin{figure*}
\centering{
\includegraphics[width=15cm]{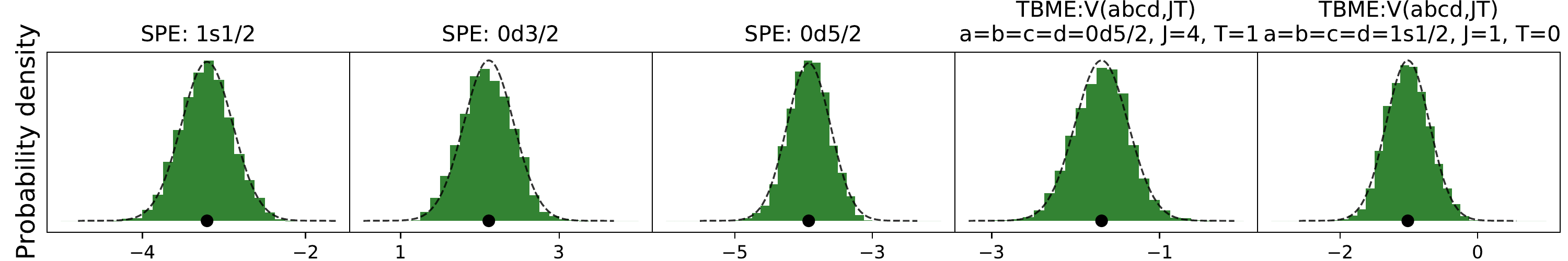}
\caption{Probability density distributions for some selected parameters. The x-axis is the parameter value in units of MeV, and the y-axis is probability density for prior (dashed line) and posterior (histogram).
The histograms are drawn by the 100,000 MCMC samples with a bin width of 125 keV. 
The first three are SPEs of $1s_{1/2}, 0d_{3/2}, 0d_{5/2}$, and the rest are the TBMEs $V(abcd;JT)$ with $(a=b=c=d=0d5/2,J=4,T=1)$ and $(a=b=c=d=1s1/2,J=1,T=0)$.
Both the histograms and Gaussians are normalized to represent probability distributions.
\label{pic:theta_hist}}}
\end{figure*}

To sample from the posterior $\propto L(\vec{c}) \times Pr(\vec{c})$, we adopted a Markov chain Monte Carlo (MCMC) method with the Metropolis-Hastings algorithm.
We set a random interaction around the USDB parameters as an initial state of the MCMC,
and generated 100,000 MCMC samples (after 10,000 burn-in) with the Metropolis-Hastings updates.
The elapsed time to achieve it with the EC emulator on a laptop was about 30 minutes.
In Fig.~\ref{pic:theta_hist}, we show one realization of parameter-distributions for some selected channels. The posteriors are drawn by the histograms with a bin width of 125 keV.
The dot symbols and dashed lines denote the USDB and the prior distributions by independent normal distributions.
The complete plot including the other TBMEs and technical details are summarized in Sec. C of Supplementary Material~\cite{supple}, and one can see that the independent runs give the converged energy distributions.

The posterior is similar to the prior, but some are slightly modified through the log-likelihood, Eq.~\eqref{eq:llh}. This indicates that the USDB values for these channels were affected and thereby constrained by a variety of data, not just the ${}^{28}$Si data with lower $J$.
In this way, the efficient emulator allows a detailed sensitivity analysis on the parameter and adopted data.

There still remain a lot of investigations undone such as dependence on the choices of the loss function (log-likelihood), experimental data, and $\sigma_{\mathrm{err},i}$, and the more efficient sampling schemes for MCMC.
In addition to these, it is also desired to develop a method to quantify the uncertainties taking into account of the variational property of the EC estimates for the energy eigenvalues.
However, the above analysis manifests that one can achieve the samplings,
which were computationally too heavy to carry out on a laptop, in tens of minutes with the help of the EC emulator.
It is encouraging to proceed towards full evaluation of uncertainties for deeper understanding of the relation among nuclear structure, effective interactions, the choice of the experimental data, and the underlying nuclear force.

\subsection{Accuracy of the approximate wave functions\label{sec:EM}}

\begin{figure}[h]
\centering{
\includegraphics[width=15cm]{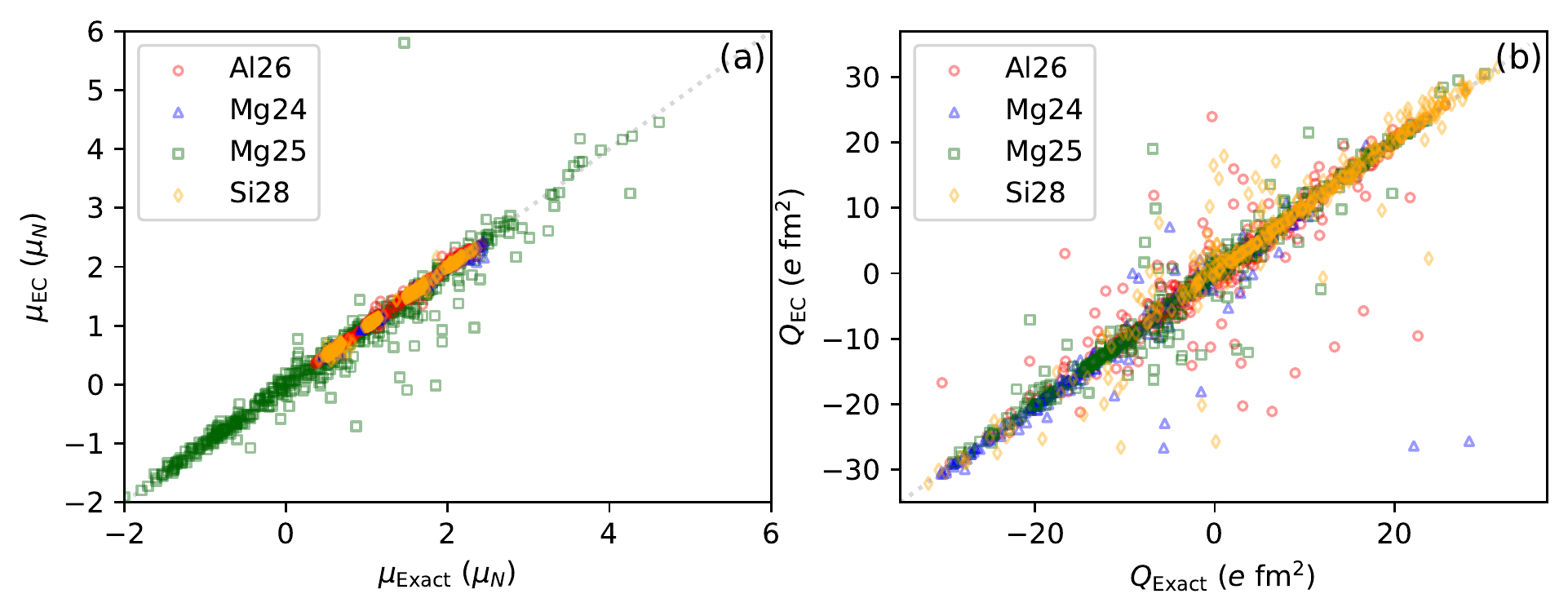}
\caption{
The EC estimates of (a) magnetic dipole moments $\mu$ in units of nuclear magneton $\mu_N=0.105$ $e$ fm and (b) quadrupole moments $Q$ in units of $e$ fm$^2$ of the yrast states of the four $sd$-shell nuclei in comparison to the exact values.
The symbols denote the yrast states with $J$ ranging from $1/2$ to $9/2$ by increment of $1/2$ which are allowed by the selection rules.
The values are evaluated with the free-nucleon values for $g$-factors and effective charges ($e_p=1.5, e_n=0.5$).
\label{pic:moment}}}
\end{figure}

Next, we test the accuracy of the approximate wave functions by evaluating other observables using Eq.~\eqref{eq:obs}.
In Fig.~\ref{pic:moment}, we show the approximated values of magnetic dipole moments $\mu$ and electric quadrupole moments $Q$, which are calculated for the validation set, of the yrast states of the four nuclei in the middle of $sd$ shell.
The symbols denote the yrast states with $J$ ranging $1/2,1,...,9/2$ which are allowed by the selection rules, i.e., $J \neq 0$ for $\mu$ and $J \neq 0,1/2$ for $Q$.
The values are given in units of nuclear magneton $\mu_N=0.105$ $e$ fm for $\mu$ and $e$ fm$^2$ for $Q$, and we used the free-nucleon values for the $g$-factors ($g_{\ell p}=1.0$, $g_{\ell n}=0$, $g_{sp} = 5.586$, and $g_{sn}=-3.826$) and effective charges ($e_p=1.5, e_n=0.5$), while these choices do not have much effect on the correlation plot.

One can see that the agreement between the EC estimates and the exact ones for both $\mu$ and $Q$ become worse than that for energy eigenvalues.
This is simply because the electromagnetic transitions are much sensitive to the accuracy of the wave functions.
Note, however, that the correspondence between the exact and approximated eigenstates are deduced simply from their energy eigenvalues.
As described below, it is highly non-trivial to find one by one correspondence among them.

Most of the approximate values of the magnetic dipole moment are nearly on the diagonal line in Fig.~\ref{pic:moment} (a).
The points for even nuclei with $J=1,2,3,4$ concentrate on $\mu\sim 0.5,1.0,1.5,2.0$, respectively.
On the other hand, results of electric quadrupole moments are rather scattered,
while most of the results are still on the diagonal line.
As an extreme example, let us look at the lower right-most triangle in Fig.~\ref{pic:moment} (b), which is $J=4^+_1$ state of ${}^{24}$Mg with $Q_\mathrm{Exact}=+28.340$ ($e$ fm$^2$). 
Diagonalizing the corresponding random interaction, we found that the first and second eigenstates have small energy difference $\sim$ 400 keV and, the $Q$-moments with opposite sign and similar amplitude. While the approximate energy eigenvalues are close to the exact ones with $\sim2\%$ accuracy, the $Q$ moment by EC has the opposite sign to the exact ones:
\begin{align}
E_\mathrm{Exact}(4^+_1) & = -75.951\ \mathrm{MeV}, \ 
Q_\mathrm{Exact} (4^+_1) =  +28.340\ e\mathrm{fm}^2, \nonumber \\
E_\mathrm{Exact}(4^+_2) &= -75.454 \ \mathrm{MeV}, \ 
Q_\mathrm{Exact} (4^+_2) =  -25.682\ e\mathrm{fm}^2, \nonumber \\
E_\mathrm{EC}(4^+_1) & =  -74.751 \ \mathrm{MeV}, \ 
Q_\mathrm{EC} (4^+_1)  = -25.635\ e\mathrm{fm}^2, \nonumber \\ 
E_\mathrm{EC}(4^+_2) &=  -73.825 \ \mathrm{MeV}, \ 
Q_\mathrm{EC} (4^+_2)  =  + 27.599\ e\mathrm{fm}^2.
\end{align}
This indicates that the approximate eigenvectors for the nearly degenerate states can be in the wrong order in terms of the energy, or contaminated by other states.

In addition to the energy eigenvalues,
the moments and electromagnetic transitions are much useful for sensitivity analyses
on the input parameters and on adopted approximations and truncations (if there were).
When discussing these observables, one needs additional efforts to improve the accuracy of the approximate wave functions.
We revisit this issue in Sec.~\ref{sec:ReEM}.

\subsection{Acceleration of the Lanczos iterations by the preprocessing\label{sec:preprocess}}

From the results shown in the previous subsections, it is expected that the convergence of the EC estimates to the exact values is not so fast as a function of the sample size.
While the eigenvector continuation provides an efficient emulator,
one still needs additional efforts to discuss the observables which demand high accuracy of the wave functions.

In general, the matrix-vector product of the shell-model Hamiltonian matrix takes up most of the calculation time, and one usually starts the Lanczos iterations from a normalized random vector.
In the following, we consider the preprocessing, which is to start the Lanczos iterations from the EC approximate eigenvectors, to accelerate the shell-model calculations.
For simplicity, we discuss the numbers of Hamiltonian operations in the Lanczos iterations with and without the preprocessing, although the elapsed time of the computation depends not only on the number of the operations but also on various conditions such as computing environment and the block size for a block algorithm.

As an example, we consider the four $sd$-shell nuclei discussed above, and fix the target shell-model Hamiltonian to the USDB interaction.
In Table~\ref{tab:iter}, we summarized the number of iterations, which is equivalent to the number of Hamiltonian operations, required to get converged results of the lowest states with the specified $J$.
The number of iterations for the block Lanczos method is summarized
in Table~\ref{tab:iter_block}.
When considering the excited states too, we use the block algorithm for the Lanczos method
and both the block size $q$ and the number of states of interest $n$ are fixed as four to simplify the discussions.
It means that we prepare the four lowest approximate eigenvectors from the samples and then use them as the starting block vectors with the size of $q=4$.

\begin{table}
\centering{
\caption{The number of Lanczos iterations required to obtain the converged results of the four nuclei in the middle of $sd$ shell.
The $\sim$ symbol means that the number can change depending on the random seed.
\label{tab:iter}}
\begin{tabular}{lllll}
\\
\hline
\hline
starting vector  &   \multicolumn{4}{l}{$\#$ iterations }\\
\cline{2-5}
 &  ${}^{28}$Si  & 
${}^{26}$Al& 
${}^{25}$Mg &
${}^{24}$Mg \\
& $(0^+_1)$ & 
$(0^+_1)$ & 
$(\frac{1}{2}^+_1)$ &
$(0^+_1)$ \\
\hline
random & $\sim 26$& $\sim 24$&  $\sim 33$ & $\sim 18$ \\
EC$(N_s=50)$ & 15  & 15 & 17& 9 \\
EC$(N_s=50^*)$  & 15 & 14 &13  & 9 \\
EC$(N_s=150)$  &13 & 11 & 10  &6  \\
EC$(N_s=250)$   & 13& 9 & 9 & 5 \\
\hline
\hline
\end{tabular}
\caption{The number of block Lanczos $(q=4)$ iterations required to obtain the converged results of four excited states with $J=0$ (even nuclei) or $J=1/2$ (odd nuclei).
\label{tab:iter_block}}
\begin{tabular}{lllll}
\\
\hline\hline
 starting vector &   \multicolumn{4}{l}{$\#$ iterations }\\
\cline{2-5}
&  ${}^{28}$Si &
${}^{26}$Al  & 
${}^{25}$Mg  &
 ${}^{24}$Mg \\
& 
$(0^+_{1-4})$ & 
$(0^+_{1-4})$ & 
$(\frac{1}{2}^+_{1-4})$ &
$(0^+_{1-4})$ \\ 
\hline
random                & $\sim30$ & $\sim 29$ & $\sim 30$  & $\sim 28$ \\
EC($N_s$=$50$)     &22  &16  & 20 & 15   \\
EC($N_s$=$50^*$)  & 22 & 19  & 21 & 16 \\
EC($N_s$=$150$)   & 22& 17 & 18 & 11 \\
EC($N_s$=$250$)   & 20 & 16 & 16 & 8 \\
\hline\hline
\end{tabular}
}
\end{table}

As a general trend, we can see that the number of iterations becomes smaller as the number of samples increases. This tendency is prominent in the case of the single target state, Table~\ref{tab:iter}.
One exception is seen in the ${}^{26}$Al in Table~\ref{tab:iter_block}; the $N_s=50^*$ and $N_s=150$ cases are worse than the $N_s=50$ case.
This suggests that in order to speed up the convergence, the initial guesses of the nearly degenerate states must be made simultaneously with high accuracy.
Indeed, the exact eigenvalues of third and fourth $J=0$ states of ${}^{26}$Al are close to each other with a difference of about $50$ keV.

It should be noted that the preprocessing can deteriorate the convergence with some naive usage.
If one calculates the ten lowest $J=0$ states of ${}^{24}$Mg starting from four approximate eigenvectors which are constructed from the $N_s=250$ $(50\times 5)$ samples,
one may need additional several iterations compared to a random starting vector.

\begin{figure}[t]
\centering{
\includegraphics[width=12cm]{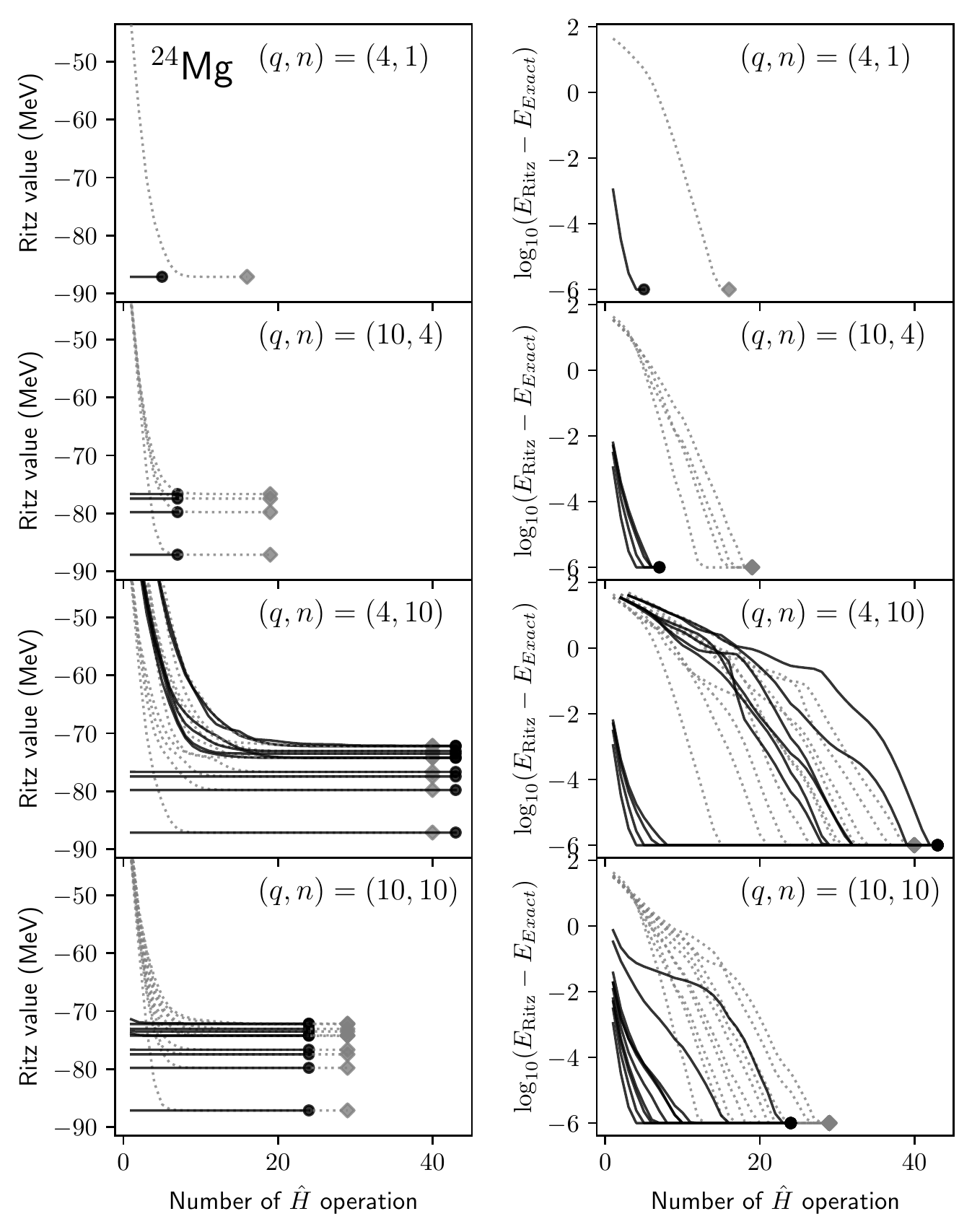}
\caption{Convergence pattern of the block Lanczos iterations for $J=0$ states of ${}^{24}$Mg.
The four rows, respectively, correspond to $(q,n)=(4,4), (10,4), (4,10), (10,10)$ cases, where $q$ is the block size and $n$ is the number of states of interest during the Lanczos iterations.
The left panels are for Ritz values, and the right panels show the difference on a logarithmic scale from the exact value with the Lanczos tolerance $10^{-6}$.
The solid lines with filled circle and dashed lines with diamond denote the cases starting from preprocessed and random initial vectors, respectively.
The symbols represent the points at which the results get converged.
\label{pic:convMg24}}}
\end{figure}

The convergence patterns of the block Lanczos iterations are shown in Fig.~\ref{pic:convMg24}.
The four rows from the top to the bottom corresponds to the case of $(q,n)= (4,4), (10,4), (4,10), (10,10)$, respectively.
Here, the $q$ is the block size and the $n$ is the number of states of interest.
The right panels are difference in a logarithmic scale from the exact value.
Note that the tolerance for the Lanczos method is set as $10^{-6}$, and the difference smaller than the tolerance is regarded as $\log_{10}(E_\mathrm{Ritz}-E_\mathrm{Exact})=-6$ in the log-plot.
The solid line with circles and the dashed lines with diamonds correspond to the cases of preprocessed and random initial vectors, respectively.
The symbols represent the points at which the results converged.

As already discussed above, the $(q,n)= (4,4)$ case exhibits the effect of the preprocessing. 
Looking at $(q,n)= (10,4)$, we can see that the number of Hamiltonian operation is much reduced.
On the other hand, the $(q,n)= (4,10)$ case shows that insufficient preprocessing can deteriorate the convergence, 
and the effect of the preprocessing is limited for the $(q,n)= (10,10)$ case due to the ninth and tenth lowest states.
This is because the $N_s=250$ $(50\times5)$ samples have less information about the higher excited states.

As shown in Fig.~\ref{pic:convMg24} and Supplementary Figure D1-D3~\cite{supple}, the efficiency of the preprocessing depends on the problem.
A rule of thumb to receive the benefit of the preprocessing is starting with a larger number of states than that of interest.
It does not always reduce the computation time, because the number of manipulations in each Lanczos iteration becomes larger as the block size increases.
In such a case, it would be better to start with a large initial block vector and, after several iterations, compress it to a smaller block vector by the Thick-Restart method.

One can consider another preprocessing by taking eigenstates within a truncated subspace (e.g., particle-hole truncation).
In general, however, the overlap between the two eigenvectors obtained within a truncated subspace and the full model space
strongly depends on the states and the choice of the truncation scheme for those target states.
For example, third $0^+$ state of ${}^{56}$Ni is known to show rather slow convergence
as a function of the number of allowed particle-hole excitations in the truncation scheme than yrast states~\cite{HoroiNi56}.
The EC emulator gives a more general way to make starting vectors for nuclei and/or states without such a special consideration.

\subsection{Improvement of approximate wavefunctions by the preprocessing \label{sec:ReEM}}

In Sec.~\ref{sec:EM}, we showed that the EC emulator can fail to approximate the magnetic dipole moments and electric quadrupole moments due to the level inversion for nearly degenerate levels.
Here, we reconsider this issue and its improvement utilizing the preprocessing discussed above.

\begin{figure}[h]
\centering{
\includegraphics[width=15cm]{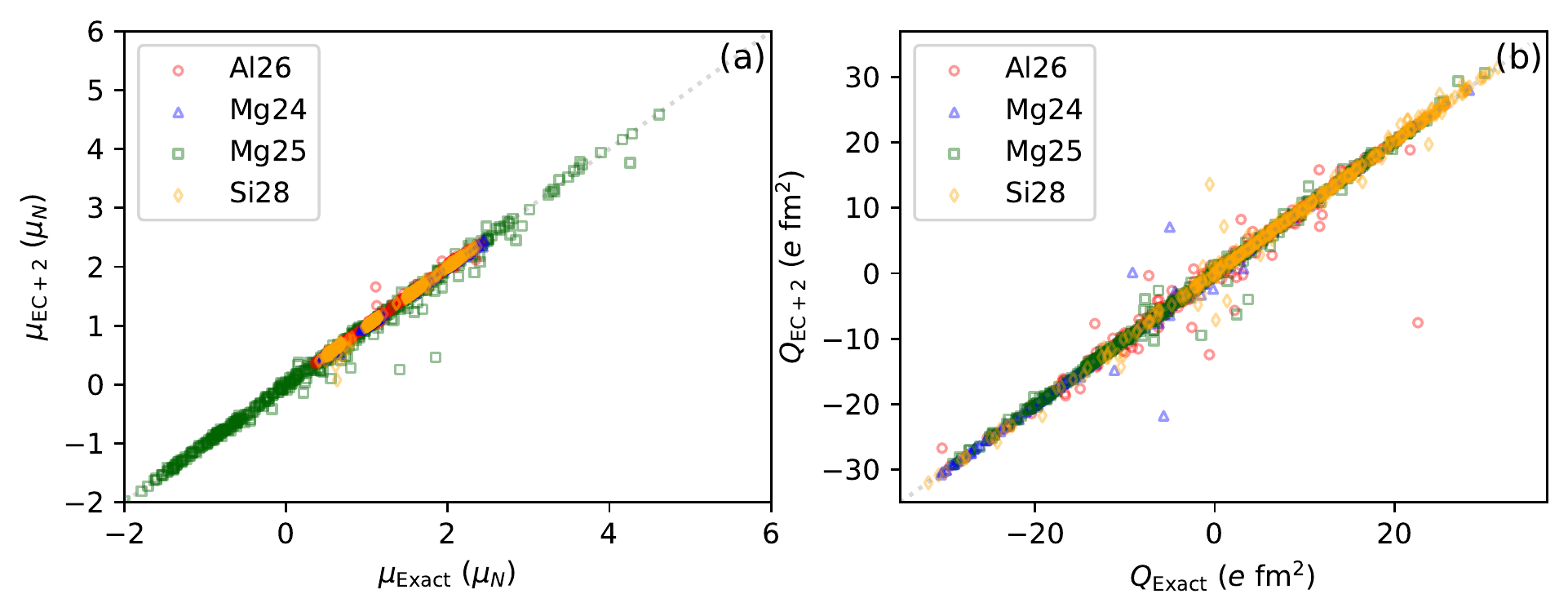}
\caption{
The EC$+2$ estimates of $\mu$ and $Q$ moments.
The symbols and lines are the same as Fig.~\ref{pic:moment}.
\label{pic:plus2_moment}}}
\end{figure}

For this purpose, taking EC emulated wave functions as the starting Lanczos vectors, we perform only two additional Lanczos iterations. We will call it EC$+2$, and the reevaluation of $\mu$ and $Q$ moments with EC$+2$ wavefunctions are shown in Fig.~\ref{pic:plus2_moment}.
It is seen that the scatter of the symbols has been considerably improved with just a few additional Lanczos iterations.
For some points that are still far from the diagonal line, these are, in general, corresponding to the random interactions that induce nearly degenerate states within a few tens of keV, so that their moments are very sensitive to the number of Lanczos iterations.

This result manifests the efficacy of our workflow using EC as an emulator and a preprocessor to estimate any observables of interest. 

\subsection{Extensibility of EC emulator: to go beyond the sd-shell\label{sec:exten}}

\begin{figure*}[h]
\centering{
\includegraphics[width=15cm]{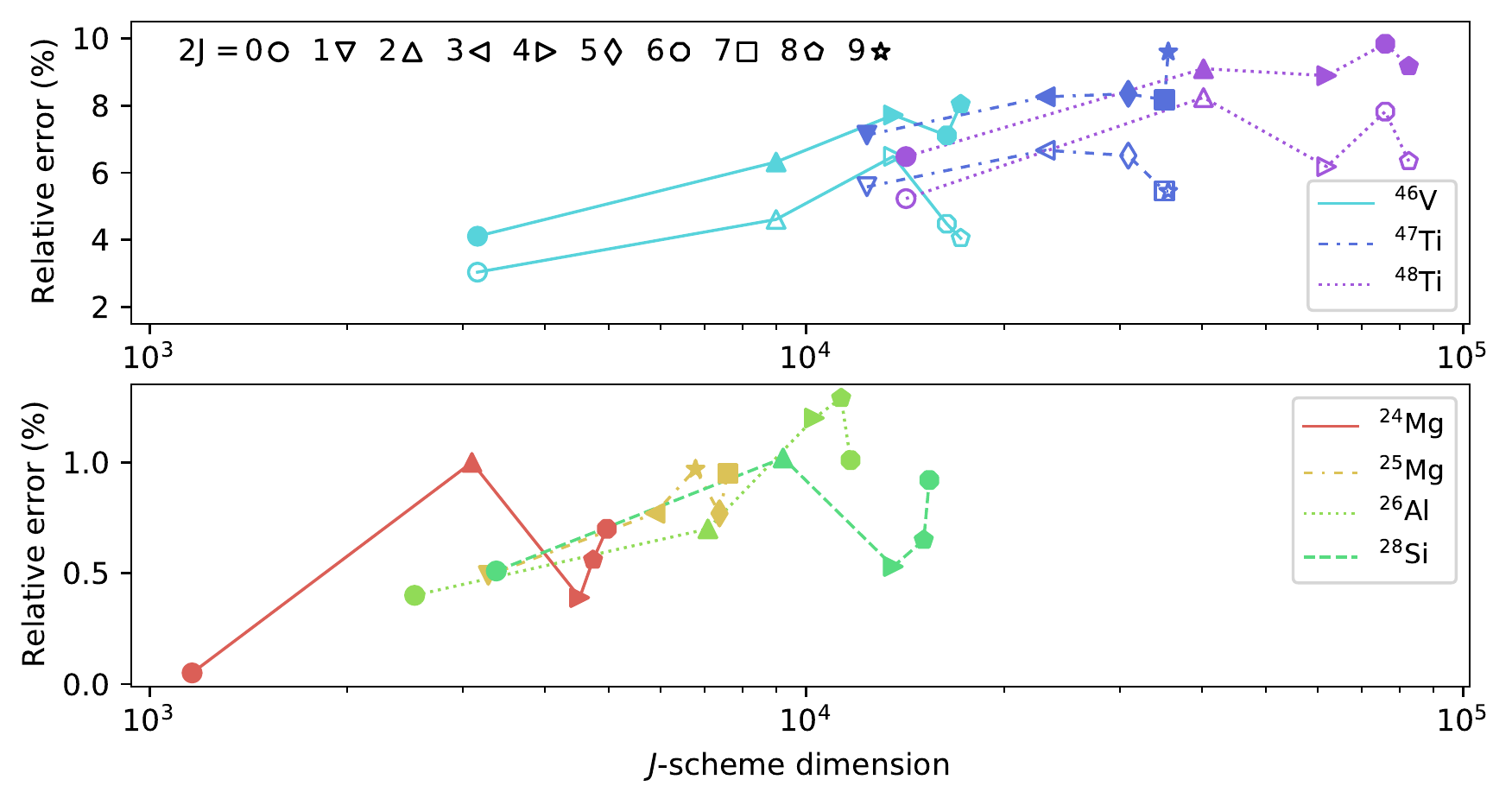}
\caption{
Relative errors with $N_s=250$ samples against $J$-scheme dimensions for ${}^{24,25}$Mg, ${}^{26}$Al,
${}^{28}$Si (lower panel), ${}^{46}$V, and ${}^{47,48}$Ti (upper panel).
The filled symbols correspond to the samples generated by varying all the parameters with $\sigma_\mathrm{int.}=1$ around the reference values (USDB and GXPF1A).
The open symbols in the upper panel show the results with samples in which only the 32 parameters related to $f7/2$ and $p3/2$ were varied with the same $\sigma_\mathrm{int.}$.
\label{pic:dimscale}}}
\end{figure*}

Regarding the extensibility of the EC emulator for shell-model calculations for larger systems,
we explore the behavior of the accuracy of the emulator against the model space size in this subsection.
In Fig.~\ref{pic:dimscale}, we plot the relative errors defined in Eq.~\eqref{eq:relerr} as a function of $J$-scheme dimension, 
\begin{align}
d_{J}(J) = d_{M}(M=J) - d_{M}(M=J+1),
\end{align}
where $d_J(J)$ and $d_M(M)$ are $J$- and $M$-scheme dimensions under given total angular momentum $J$ and its projection $M$, respectively.
It makes dependence of the accuracy on the dimension and total $J$ clearer.
As a whole, the relative errors are increasing functions of $J$-scheme dimension,
and the lower $J$ states show relatively faster convergence to the exact ones.
Since the smaller $J$ generally gives smaller $J$-scheme dimension,
the results indicate that the number of effective degrees of freedom governs the accuracy.
From Fig.~\ref{pic:dimscale}, we can understand the reason why the excitation energies show slower convergence to the exact one compared to the absolute values, as already shown in Tab~\ref{tab:err}.

In addition to the four $sd$-shell nuclei discussed above, we show the results with lower $pf$-shell nuclei (${}^{46}$V, and ${}^{47,48}$Ti) to see the model-space dependence.
For the $pf$-shell model space, there are 199 parameters (4 SPEs and 195 TBMEs) in isospin formalism, which is three-times larger than the $sd$-shell case, and we used the GXPF1A interaction~\cite{GXPF1A} as the reference to prepare the random samples.
The other steps such as generating samples are the same as in the case of the $sd$ shell.
As a whole, the relative errors become worse than the $sd$-shell case, when we use $\sigma_\mathrm{int.}=1$ for 199 parameters.
For the target nuclei (${}^{46}$V, and ${}^{47,48}$Ti), it seems that the SPEs and TBMEs involved in $f7/2$ and $p3/2$ mainly affect the wavefunctions, so random samples made in such a way that all the parameters are independent can be less informative to span the exact eigenvectors, and thereby gives larger errors than the $sd$-shell case.
Indeed, if we restrict the sampling space to the parameters related only to $0f7/2$ and $1p3/2$ (i.e. the other parameters are fixed as the reference values), the relative errors against the validation set are improved especially for higher $J$ states. The corresponding results are shown by the open symbols in Fig.~\ref{pic:dimscale}.

For larger model spaces, it is expected that a strategy such as principal component analysis to sample mainly in low-dimensional subspaces that are sensitive to wavefunctions will be necessary to achieve the same level of accuracy as in the $sd$-shell case.
Additionally, if one goes beyond the $sd$ shell and lower $pf$ shell,
we need code development with massive parallelization and more efficient designs of the sampling scheme.
Even in such cases, the workflow re-using already sampled wave functions
gives acceleration cycles for the sampling procedure, as demonstrated in the previous subsections.

\section{Summary and Outlook}\label{sec:Summary}

We demonstrated that the eigenvector continuation (EC) can be used as an efficient emulator for shell-model calculations, as in previous works for other models~\cite{EC_CC,EC_Scattering,EC_Scattering2,EC_Scattering3,EC_NCSM,EC_RMat,EC_LEC_NCSM}.
As an example, we considered $sd$-shell nuclei and the randomly sampled effective interactions around the USDB interaction.

With the sample eigenvectors under the 50 random interactions,
we demonstrated that the exact eigenenergies under other 100 random interactions in a similar range can be estimated with   an accuracy of a few percent.
The EC emulator is found to be helpful for rough parameter optimization and quantifying uncertainty.
The latter was partially demonstrated by Markov chain Monte Carlo sampling on a simplified problem.

It should be noted that one should pay attention to the accuracy of the approximate wave functions by EC when discussing excitation energies and observables such as electromagnetic transitions,
which are sensitive to the relative convergence speed among different states.
In such cases, one needs additional efforts to improve the accuracy of the approximate wave functions.
As one solution to make it feasible, we proposed a new usage of the eigenvector continuation
as the preprocessing for the shell-model calculations.
The exact diagonalization and subsequent manipulations can be accelerated by starting the Lanczos iterations from the EC approximate eigenvectors.

As is schematically shown in Fig.~\ref{pic:flowchart},
the eigenvector continuation can be used (a) as an efficient emulator, (b) as a preprocessing to accelerate the exact calculations, and (c) to generate another sample eigenvector efficiently with the help of the preprocessing.
The eigenvectors in the optimization process of effective interactions are no longer necessarily to be discarded, even if their eigenvalues are far from the corresponding experimental data.
These eigenvectors are not wasted, but can become accumulated knowledge and accelerate the workflow of researchers.

In addition to the methodology, we developed the new open-source shell-model code, ShellModel.jl, written in the Julia language.
The code enables us to sample shell-model results efficiently, and is highly flexible to add any extensions to suit the users' purposes such as MCMC samplings.
The main codes and sample scripts are provided in the GitHub repository~\cite{SY_Github}.

There are many possible future works along the line presented in this work.
While the microscopically derived effective interactions give reasonable results,
it is still indispensable to construct {\it better} phenomenological interactions for
more quantitative discussions in comparison with experimental data,
designing new experiments, and testing the validity of many-body methods (and the approximations in them) 
to derive effective interactions.
Of particular importance for this purpose is the extension of the proposed method to larger model spaces.
This requires a more efficient sampling strategy that can work even when the computational cost for a single sample is high.
Such a development is expected to facilitate the construction of phenomenological interactions and the validation of microscopically derived effective interactions, even in mass regions where gold-standard phenomenological interactions like USDB do not exist.

For parameter calibration using the EC emulators, we still have interesting topics to be further studied.
Unlike phenomenological interactions used in this study,
the dependence of the wavefunctions on the controlling parameters such as low-energy constants (LECs) in chiral EFT
will not be factorized out as simple as Eq.~\eqref{eq:TDfac}
when a shell model effective interaction derived from a chiral potential is considered.
In such a case, sensitivity analyses to the parameters will become more beneficial, but more non-trivial.
If one could extend the EC emulator to treat the LECs as the controlling parameters for shell-model effective interactions,
uncertainties in the EC emulator would be reduced.
This is because the number of parameters is reduced from e.g., 199 ($pf$-shell case above) to the typical number of LECs, 30--40.
Those sensitivity analyses on the LECs in medium mass regions will make
the relationship between nuclear structure and nuclear force clearer.

For uncertainty quantification, efficient emulators allow to evaluate full posteriors for the parameters and/or observables.
It would also be interesting to explore such a direction using more efficient MCMC methods.

It is also desirable to open a database of transition densities and wave functions so that one can easily obtain approximate eigenpairs and benefit from the preprocessing without repeating massive computations. This will further facilitate the cycle of cross-validation between theory and experiment.
These would contribute to the long journey towards understanding of how nuclei emerge from the underlying nuclear force.

\section*{Acknowledgments}

This work was partly supported by KAKENHI grants (17K05433) from JSPS, ``Priority Issue on
post-K computer" (Elucidation of the Fundamental Laws and
Evolution of the Universe), ``Program for Promoting Researches on the Supercomputer Fugaku" (JPMXP1020200105), MEXT, Japan, the Research Project Promotion Grant for Young Researchers of Utsunomiya University.
We also thank Multidisciplinary Cooperative Research Program
by Center for Computational Sciences, Tsukuba University (xg18i035).

\vspace{0.2cm}
\noindent


\bibliography{2111-019-3D-SotaYoshida}

\begin{thebibliography}{10}

\bibitem{Brown_Rev}
B~A Brown, Progress in Particle and Nuclear Physics, {\bf 47}(2), 517 -- 599
  (2001).

\bibitem{CaurierRev}
E.~Caurier, G.~Mart\'{\i}nez-Pinedo, F.~Nowacki, A.~Poves, and A.~P. Zuker,
  Rev. Mod. Phys., {\bf 77}, 427--488 (Jun 2005).

\bibitem{RMP_Otsuka}
Takaharu Otsuka, Alexandra Gade, Olivier Sorlin, Toshio Suzuki, and Yutaka
  Utsuno, Rev. Mod. Phys., {\bf 92}, 015002 (Mar 2020).

\bibitem{EGMrev}
E.~Epelbaum, H.-W. Hammer, and Ulf-G. Mei\ss{}ner, Rev. Mod. Phys., {\bf 81},
  1773--1825 (Dec 2009).

\bibitem{EMrev}
R.~Machleidt and D.R. Entem, Phys. Rept., {\bf 503}(1), 1 -- 75 (2011).

\bibitem{VS-IMSRG1}
S.~K. Bogner, H.~Hergert, J.~D. Holt, A.~Schwenk, S.~Binder, A.~Calci,
  J.~Langhammer, and R.~Roth, Phys. Rev. Lett., {\bf 113}, 142501 (Oct 2014).

\bibitem{VS-IMSRG2}
S.~R. Stroberg, H.~Hergert, J.~D. Holt, S.~K. Bogner, and A.~Schwenk, Phys.
  Rev. C, {\bf 93}, 051301 (May 2016).

\bibitem{VS-IMSRG3}
S.~R. Stroberg, A.~Calci, H.~Hergert, J.~D. Holt, S.~K. Bogner, R.~Roth, and
  A.~Schwenk, Phys. Rev. Lett., {\bf 118}, 032502 (Jan 2017).

\bibitem{Stroberg_rev19}
S.~Ragnar Stroberg, Heiko Hergert, Scott~K. Bogner, and Jason~D. Holt, Annual
  Review of Nuclear and Particle Science, {\bf 69}(1), 307--362 (2019).

\bibitem{VS-IMSRG_Miyagi}
T.~Miyagi, S.~R. Stroberg, J.~D. Holt, and N.~Shimizu, Phys. Rev. C, {\bf 102},
  034320 (Sep 2020).

\bibitem{Stroberg2021}
S.~R. Stroberg, J.~D. Holt, A.~Schwenk, and J.~Simonis, Phys. Rev. Lett., {\bf
  126}, 022501 (Jan 2021).

\bibitem{CCEI1}
G.~R. Jansen, J.~Engel, G.~Hagen, P.~Navratil, and A.~Signoracci, Phys. Rev.
  Lett., {\bf 113}, 142502 (Oct 2014).

\bibitem{CCEI2}
G.~R. Jansen, M.~D. Schuster, A.~Signoracci, G.~Hagen, and P.~Navr\'atil, Phys.
  Rev. C, {\bf 94}, 011301 (Jul 2016).

\bibitem{EI_NCSM1}
E.~Dikmen, A.~F. Lisetskiy, B.~R. Barrett, P.~Maris, A.~M. Shirokov, and J.~P.
  Vary, Phys. Rev. C, {\bf 91}, 064301 (Jun 2015).

\bibitem{EI_NCSM2}
N.~A. Smirnova, B.~R. Barrett, Y.~Kim, I.~J. Shin, A.~M. Shirokov, E.~Dikmen,
  P.~Maris, and J.~P. Vary, Phys. Rev. C, {\bf 100}, 054329 (Nov 2019).

\bibitem{SMCC1}
Z.~H. Sun, T.~D. Morris, G.~Hagen, G.~R. Jansen, and T.~Papenbrock, Phys. Rev.
  C, {\bf 98}, 054320 (Nov 2018).

\bibitem{SMCC2}
Z.~H. Sun, G.~Hagen, G.~R. Jansen, and T.~Papenbrock, Phys. Rev. C, {\bf 104},
  064310 (Dec 2021).

\bibitem{Carlsson2016PRX}
B.~D. Carlsson, A.~Ekstr\"om, C.~Forss\'en, D.~Fahlin Str\"omberg, G.~R.
  Jansen, O.~Lilja, M.~Lindby, B.~A. Mattsson, and K.~A. Wendt, Phys. Rev. X,
  {\bf 6}, 011019 (Feb 2016).

\bibitem{Melendez2017}
J.~A. Melendez, S.~Wesolowski, and R.~J. Furnstahl, Phys. Rev. C, {\bf 96},
  024003 (Aug 2017).

\bibitem{Wesolowski2019}
S~Wesolowski, R~J Furnstahl, J~A Melendez, and D~R Phillips, Journal of Physics
  G: Nuclear and Particle Physics, {\bf 46}(4), 045102 (feb 2019).

\bibitem{UQ_DFT}
J.~D. McDonnell, N.~Schunck, D.~Higdon, J.~Sarich, S.~M. Wild, and
  W.~Nazarewicz, Phys. Rev. Lett., {\bf 114}, 122501 (Mar 2015).

\bibitem{UQ_LDM}
Bartholom\'e Cauchois, Hongliang L\"u, David Boilley, and Guy Royer, Phys. Rev.
  C, {\bf 98}, 024305 (Aug 2018).

\bibitem{UQ_NM_1}
C.~Drischler, J.~A. Melendez, R.~J. Furnstahl, and D.~R. Phillips, Phys. Rev.
  C, {\bf 102}, 054315 (Nov 2020).

\bibitem{UQ_NM_2}
C.~Drischler, R.~J. Furnstahl, J.~A. Melendez, and D.~R. Phillips, Phys. Rev.
  Lett., {\bf 125}, 202702 (Nov 2020).

\bibitem{UQ_reaction}
G.~B. King, A.~E. Lovell, L.~Neufcourt, and F.~M. Nunes, Phys. Rev. Lett., {\bf
  122}, 232502 (Jun 2019).

\bibitem{SY_UQ}
Sota Yoshida, Noritaka Shimizu, Tomoaki Togashi, and Takaharu Otsuka, Phys.
  Rev. C, {\bf 98}, 061301 (Dec 2018).

\bibitem{Fox_UQ}
Jordan M.~R. Fox, Calvin~W. Johnson, and Rodrigo~Navarro Perez, Phys. Rev. C,
  {\bf 101}, 054308 (May 2020).

\bibitem{supple}
Supplementary material,
\newblock URL will be inserted.

\bibitem{USDB}
B~A Brown and W~A Richter, Phys. Rev. C, {\bf 74}, 034315 (Sep 2006).

\bibitem{KB3G}
A.~Poves, J.~Sánchez-Solano, E.~Caurier, and F.~Nowacki, Nuclear Physics A,
  {\bf 694}(1), 157--198 (2001).

\bibitem{GXPF}
M.~Honma, T.~Otsuka, B.~A. Brown, and T.~Mizusaki, Phys. Rev. C, {\bf 65},
  061301 (May 2002).

\bibitem{GXPF1}
M.~Honma, T.~Otsuka, B.~A. Brown, and T.~Mizusaki, Phys. Rev. C, {\bf 69},
  034335 (Mar 2004).

\bibitem{GXPF1A}
{Honma, M.}, {Otsuka, T.}, {Brown, B. A.}, and {Mizusaki, T.}, Eur. Phys. J. A,
  {\bf 25}, 499--502 (2005).

\bibitem{JUN45}
M.~Honma, T.~Otsuka, T.~Mizusaki, and M.~Hjorth-Jensen, Phys. Rev. C, {\bf 80},
  064323 (Dec 2009).

\bibitem{USDI}
A.~Magilligan and B.~A. Brown, Phys. Rev. C, {\bf 101}, 064312 (Jun 2020).

\bibitem{Lanczos}
Cornelius Lanczos, J. Research Nat. Bur. Standards, {\bf 45}, 255--282 (1950).

\bibitem{NuShellX}
B.A. Brown and W.D.M. Rae, Nuclear Data Sheets, {\bf 120}, 115 -- 118 (2014).

\bibitem{BIGSTICK}
Calvin~W. Johnson, W.~Erich Ormand, and Plamen~G. Krastev, Computer Physics
  Communications, {\bf 184}(12), 2761--2774 (2013).

\bibitem{BIGSTICK2}
Calvin~W. Johnson, W.~Erich Ormand, Kenneth~S. McElvain, and Hongzhang Shan,
\newblock Bigstick: A flexible configuration-interaction shell-model code
  (2018),  {{arXiv:1801.0843}}.

\bibitem{ANTOINE}
E.~{Caurier} and F.~{Nowacki}, Acta Physica Polonica B, {\bf 30}, 705 (March
  1999).

\bibitem{MFDn1}
J.~P. Vary,
\newblock The many-fermion dynamics shell-model code,
\newblock unpublished (1992).

\bibitem{MFDn2}
J.~P. Vary and D.~C. Zheng,
\newblock The many-fermion dynamics shell-model code,
\newblock unpublished (1994).

\bibitem{MFDn3}
Philip Sternberg, Esmond~G. Ng, Chao Yang, Pieter Maris, James~P. Vary, Masha
  Sosonkina, and Hung~Viet Le,
\newblock Accelerating configuration interaction calculations for nuclear
  structure,
\newblock In {\em Proceedings of the 2008 ACM/IEEE Conference on
  Supercomputing}, SC '08, pages 15:1--15:12, Piscataway, NJ, USA (2008). IEEE
  Press.

\bibitem{MSHELL64}
Takahiro Mizusaki, Noritaka Shimizu, Yutaka Utsuno, and Michio Honma,
\newblock Mshell64 code,
\newblock unpublished.

\bibitem{KSHELL1}
Noritaka Shimizu,
\newblock Nuclear shell-model code for massive parallel computation, "kshell"
  (2013),  {{arXiv:1310.5431}}.

\bibitem{KSHELL2}
Noritaka Shimizu, Takahiro Mizusaki, Yutaka Utsuno, and Yusuke Tsunoda,
  Computer Physics Communications, {\bf 244}, 372 -- 384 (2019).

\bibitem{SY_Github}
Sota Yoshida,
\newblock Shellmodel.jl (2021),
\newblock \url{https://github.com/SotaYoshida/ShellModel.jl}.

\bibitem{Julia}
Jeff Bezanson, Stefan Karpinski, Viral~B. Shah, and Alan Edelman,
\newblock Julia: A fast dynamic language for technical computing (2012),
  {{arXiv:1209.5145}}.

\bibitem{Julia2}
Julia language~\url{https://julialang.org}.

\bibitem{Mizusaki2010}
Takahiro Mizusaki, Kazunari Kaneko, Michio Honma, and Tetsuya Sakurai, Phys.
  Rev. C, {\bf 82}, 024310 (Aug 2010).

\bibitem{EC_Frame}
Dillon Frame, Rongzheng He, Ilse Ipsen, Daniel Lee, Dean Lee, and Ermal Rrapaj,
  Phys. Rev. Lett., {\bf 121}, 032501 (Jul 2018).

\bibitem{EC_CC}
Andreas Ekstr\"om and Gaute Hagen, Phys. Rev. Lett., {\bf 123}, 252501 (Dec
  2019).

\bibitem{EC_Scattering}
R.J. Furnstahl, A.J. Garcia, P.J. Millican, and Xilin Zhang, Physics Letters B,
  {\bf 809}, 135719 (2020).

\bibitem{EC_Scattering2}
J.A. Melendez, C.~Drischler, A.J. Garcia, R.J. Furnstahl, and Xilin Zhang,
  Physics Letters B, {\bf 821}, 136608 (2021).

\bibitem{EC_Scattering3}
C.~Drischler, M.~Quinonez, P.G. Giuliani, A.E. Lovell, and F.M. Nunes, Physics
  Letters B, {\bf 823}, 136777 (2021).

\bibitem{EC_NCSM}
S.~König, A.~Ekström, K.~Hebeler, D.~Lee, and A.~Schwenk, Physics Letters B,
  {\bf 810}, 135814 (2020).

\bibitem{EC_RMat}
Dong Bai and Zhongzhou Ren, Phys. Rev. C, {\bf 103}, 014612 (Jan 2021).

\bibitem{EC_LEC_NCSM}
S.~Wesolowski, I.~Svensson, A.~Ekstr\"om, C.~Forss\'en, R.~J. Furnstahl, J.~A.
  Melendez, and D.~R. Phillips, Phys. Rev. C, {\bf 104}, 064001 (Dec 2021).

\bibitem{EC_BMBPT1}
P.~Demol, T.~Duguet, A.~Ekstr\"om, M.~Frosini, K.~Hebeler, S.~K\"onig, D.~Lee,
  A.~Schwenk, V.~Som\`a, and A.~Tichai, Phys. Rev. C, {\bf 101}, 041302 (Apr
  2020).

\bibitem{EC_BMBPT2}
P.~Demol, M.~Frosini, A.~Tichai, V.~Somà, and T.~Duguet, Annals of Physics,
  {\bf 424}, 168358 (2021).

\bibitem{EC_conv}
Avik Sarkar and Dean Lee, Phys. Rev. Lett., {\bf 126}, 032501 (Jan 2021).

\bibitem{LHS}
M.~D. McKay, R.~J. Beckman, and W.~J. Conover, Technometrics, {\bf 21}(2),
  239--245 (1979).

\bibitem{MortenG}
Morten Hjorth-Jensen, Thomas~T.S. Kuo, and Eivind Osnes, Physics Reports, {\bf
  261}(3), 125--270 (1995).

\bibitem{HoroiNi56}
M.~Horoi, B.~A. Brown, T.~Otsuka, M.~Honma, and T.~Mizusaki, Phys. Rev. C, {\bf
  73}, 061305 (Jun 2006).

\end{thebibliography}

\end{document}